\documentclass[secnumarabic,amssymb,nobibnotes,aps,prl,preprint,letterpaper,preprintnumbers,amsmath,superscriptaddress,floatfix]{revtex4-1}

\usepackage{graphicx}
\usepackage[dvipsnames]{xcolor}
\usepackage{soul}
\usepackage{amsmath,bm}
\usepackage{amssymb}
\usepackage{upgreek}

\setcitestyle{super}
\makeatletter
\def\blfootnote{\xdef\@thefnmark{}\@footnotetext}
\makeatother
\usepackage{multibib}
\usepackage{natbib} 
\usepackage{etoolbox}

\usepackage{enumitem}

\usepackage{bibentry}
\nobibliography* 

\setcitestyle{super}

\renewcommand{\figurename}{\textbf{Figure}}

\begin{document}

\title{Observing Differential Spin Currents by Resonant Inelastic X-ray Scattering}

\author{Yanhong Gu$\,^{\dagger}$\blfootnote{$^{\dagger}\,$gyhshan@gmail.com, Present address: State Key Laboratory of Low Dimensional Quantum Physics, Beijing Tsinghua Institute for Frontier Interdisciplinary Innovation, Beijing 102202, China.}}
\affiliation{National Synchrotron Light Source II, Brookhaven National Laboratory, Upton, New York 11973, USA}

\author{Joseph Barker}
\affiliation{Bragg Centre for Materials Research, University of Leeds, Leeds LS2 9JT, United Kingdom}
\affiliation{School of Physics and Astronomy, University of Leeds, Leeds LS2 9JT, United Kingdom}
\affiliation{Institute for Materials Research, Tohoku University, Sendai 980-8577, Japan}

\author{Jiemin Li}
\affiliation{National Synchrotron Light Source II, Brookhaven National Laboratory, Upton, New York 11973, USA}

\author{Takashi Kikkawa}
\affiliation{Advanced Science Research Center, Japan Atomic Energy Agency, 2-4 Shirakata, Tokai-mura, Naka-gun, Ibaraki, Japan 319-1195}

\author{Fernando Camino}
\affiliation{Center for Functional Nanomaterials, Brookhaven National Laboratory, Upton, New York 11973, USA
}
\author{Kim Kisslinger}
\affiliation{Center for Functional Nanomaterials, Brookhaven National Laboratory, Upton, New York 11973, USA
}

\author{John Sinsheimer}
\affiliation{National Synchrotron Light Source II, Brookhaven National Laboratory, Upton, New York 11973, USA}

\author{Lukas Lienhard}
\affiliation{National Synchrotron Light Source II, Brookhaven National Laboratory, Upton, New York 11973, USA}

\author{Jackson J. Bauer}
\affiliation{Department of Materials Science and Engineering, Massachusetts Institute of Technology, Cambridge,
Massachusetts 02139, USA
}

\author{Caroline A. Ross}
\affiliation{Department of Materials Science and Engineering, Massachusetts Institute of Technology, Cambridge,
Massachusetts 02139, USA
}

\author{Dmitri N. Basov}
\affiliation{Department of Physics, Columbia University, New York, New York 10027, USA.}

\author{Eiji Saitoh}
\affiliation{Department of Applied Physics, The University of Tokyo, Tokyo 113-8656, Japan} 
\affiliation{Institute for AI and Beyond, The University of Tokyo, Tokyo 113-8656, Japan}
\affiliation{WPI Advanced Institute for Materials Research, Tohoku University, Sendai 980-8577, Japan} 
\affiliation{RIKEN Center for Emergent Matter Science (CEMS), Wako 351-0198, Japan}

\author{Jonathan Pelliciari}
\affiliation{National Synchrotron Light Source II, Brookhaven National Laboratory, Upton, New York 11973, USA}

\author{Gerrit E. W. Bauer}
\affiliation{Institute for Materials Research, Tohoku University, Sendai 980-8577, Japan}
\affiliation{WPI Advanced Institute for Materials Research, Tohoku University, Sendai 980-8577, Japan}

\author{Valentina Bisogni$\,^{*}$\blfootnote{$^{*}\,$bisogni@bnl.gov}}
\affiliation{National Synchrotron Light Source II, Brookhaven National Laboratory, Upton, New York 11973, USA}

\begin{abstract}
\textbf{ Controlling spin currents, i.e., the flow of spin angular momentum, in small magnetic devices is the principal objective of spin electronics, a main contender for future energy efficient information technologies. Surprisingly, a pure spin current has never been measured directly since the associated electric stray fields and/or shifts in the non-equilibrium spin-dependent distribution functions are too small for conventional experimental detection methods optimized for charge transport. Here we report that resonant inelastic x-ray scattering (RIXS) can bridge this gap by measuring the spin current carried by magnons -- the quanta of the spin wave excitations of the magnetic order -- in the presence of temperature gradients across a magnetic insulator. This is possible due to the sensitivity of the momentum- and energy-resolved RIXS intensity to minute changes in the magnon distribution under non-equilibrium conditions. We use the Boltzmann equation in the relaxation time approximation to extract transport parameters, such as the magnon lifetime at finite momentum, essential for the realization of magnon spintronics.}

\end{abstract}

\flushbottom
\maketitle

\thispagestyle{empty}


Exotic forms of transport in quantum materials can be superior to conventional charge-based transport by being faster, more energy efficient, and enabling smaller devices \cite{han2020spin,gish2024van,qian2021phonon,franchini2021polarons,choi2023nature}. Spintronics has been the vanguard of the quest to find alternative technologies to conventional electronics by controlling the flow of the intrinsic angular momentum of the 
spin \cite{current1971dyakonov,uchida2008observation,bauer2012spin}. However, a direct experimental observation of these spin currents is challenging. While non-equilibrium spin accumulations betray the detection of spin currents by readily measurable stray magnetic fields, the electric fields associated with the flow of spins\cite{hirsch1999} have never been measured. We can only deduce the existence of spin currents via the voltages caused by extrinsic spin-charge conversion processes \cite{saitoh2006conversion}.

Spin waves, the elementary excitations of the magnetic order parameter, and their quanta called magnons, are promising candidates in the search for alternative information carriers \cite{rezende2020fundamentals} studied by the field of magnonics \cite{pirro2021advances}. In high-quality magnetic materials, propagating magnons convey energy and spin angular momentum over macroscopic distances without moving charges, thereby suppressing the Joule heating. Presently, spin currents are \textcolor{black}{mostly} measured indirectly by injecting them into heavy metals that convert the current into a voltage by the inverse spin Hall effect (ISHE) \cite{saitoh2006conversion,uchida2008observation,bauer2012spin}, or directly by tracking the flow of the spin angular momentum using x-ray magnetic circular dichroism (XMCD) \cite{PKukreja2015prl, Li2016prl}. 
Here we report the first direct observation of a spin current \textcolor{black}{with energy and momentum resolution} 
by measuring resonant inelastic x-ray scattering (RIXS) spectra of the magnetic insulator yttrium iron garnet Y$_3$Fe$_5$O$_{12}$ (YIG). 

Yttrium iron garnet is an electrically insulating ferrimagnet and the material of choice for magnonics owing to its record magnetic quality \cite{serga2010yig}. A temperature gradient induces a magnon spin current from the high to the low-temperature side by the spin Seebeck effect (SSE) \cite{uchida2008observation,Uchida_2014}. However, the indirect detection of the spin current by the ISHE in platinum contacts prevents the identification of the modes responsible for the bulk transport and the role of the interface  \cite{kehlberger2015length,guo2016influence}. 

In this work we use RIXS to study the magnon spin current. 
The RIXS cross-section 
is sensitive to multiple quasi-particle excitations \cite{Ament_rixs_review, Haverkort2010}
over the whole Brillouin zone and across a large energy range, offering several advantages over other spectroscopies used to study magnons. For example, magneto-optical Kerr rotation and Brillouin light-scattering detect coherent excitations at long-wave lengths and ultra-low (GHz) frequencies \cite{Olsson2020,mclaughlin2017optical}, thereby missing out on thermal (THz) magnons that contribute to the transport by their higher density of states and group velocities \cite{chumak2015magnon}. The RIXS spectra of YIG allow us to identify the optical and acoustic magnon modes characteristic of a ferrimagnet, up to $\sim$ 100 meV.  Under non-equilibrium conditions, the applied temperature gradients modulate the scattering intensity of the acoustic magnons, even at large momentum $\mathbf{q}$ and with an energy of $\sim 10$~meV. The characteristic variation of the RIXS intensity detected under $\pm\mathbf{q}$ momentum reversal unequivocally proves the observation of the spin current. This data allows us to extract a momentum-resolved magnon transport relaxation time $\tau_{\mathbf{q}}$ = 58 ns. This 
material parameter is essential for calculating macroscopic transport properties, but so far its value has been elusive to other experimental probes.

\section*{Magnons in Yttrium Iron Garnet}

RIXS is a powerful probe for magnons and their dispersion ($\hbar\omega$ versus $\mathbf{q}$) in bulk and thin-film materials \cite{pelliciari2021tuning,gu2022site}. The magnetic scattering cross section is a product of a local and a dynamic structure factor \cite{Bisogni2014prl,jia2014persistent, Haverkort2010,Robarts2021spin_susceptibility}. The latter depends on the magnon distribution $f(\mathbf{q},\omega_\mathbf{q})$ at a given momentum transfer $\mathbf{q}$ and energy $\hbar\omega_\mathbf{q}$. The magnetic RIXS cross-section is related to the magnon distribution function  \cite{jia2016using}
\begin{equation}
\frac{d^2\sigma}{d\Omega d\hbar\omega} \sim \sum_{\mathbf{q}} \left[ \langle f(\mathbf{q} ) + 1\rangle \delta(\mathbf{Q}- \mathbf{G} - \mathbf{q}) \delta(\hbar\omega - \hbar\omega_{\mathbf{q}}) + \langle f(\mathbf{q} ) \rangle \delta(\mathbf{Q}- \mathbf{G} + \mathbf{q}) \delta(\hbar\omega + \hbar\omega_{\mathbf{q}}) \right],
\label{eqn:crosssection}
\end{equation}
where $\mathbf{Q}= \mathbf{G} ~ \textcolor{black}{\pm} ~ \mathbf{q}$ is the total momentum transfer, $\mathbf{G}$ is a reciprocal-lattice vector and \(\hbar\omega=\hbar\omega_{\rm{i}}-\hbar\omega_{\rm {f}}\) is the difference between incoming and scattered photon energies. Vectors (scalars) are expressed with bold (regular) font. 
The first term in Eq.~\ref{eqn:crosssection} with \(\hbar\omega>0\) represents magnon creation and photon energy loss (Stokes channel), while the second term with \(\hbar\omega<0\) refers to annihilation of thermally generated magnons and photon energy gain (anti-Stokes channel). The precise proportionality of the terms in Eq.~\ref{eqn:crosssection} has not been extensively discussed for RIXS. 
 
The RIXS intensity of YIG is resonantly enhanced at the x-ray absorption edges of the Fe atoms in the two inequivalent sites, i.e. the tetrahedral ($d$-site) and octahedral ($a$-site) oxygen cages (see unit cell in Fig.~\ref{fig:1_1}\textbf{a}). The Fe L$_{3,2}$-edge x-ray absorption spectroscopy (XAS) in the XMCD mode is highly sensitive to the spin orientation and chemical environment  \cite{Vasili2017XMCD}. 
Figure~\ref{fig:1_1}\textbf{b} shows the room temperature x-ray absorption spectra (XAS) 
measured with positive $(C^+)$ and negative $(C^-)$ circular light polarization (see Methods for details). The XMCD signal ($I^--I^+$) (blue line) agrees with previous measurements \cite{Vasili2017XMCD}. The negative XMCD response at $\sim710$~eV comes from the majority tetrahedral $d$-sites and corresponds to the transition from filled Fe 2$p_{3/2}$ core levels to unoccupied Fe 3$d_{e}$ levels. 
For RIXS, we tune the resonant energy to $\sim710$~eV, as it maximizes the sensitivity to magnetic effects
\cite{elnaggar2019magnetic}. 

Figure~\ref{fig:1_1}\textbf{c} shows the RIXS spectrum of a YIG single crystal at $\mathbf{q}$ = [0.2, 0.2, 0.2] in reciprocal lattice units (r.l.u.), at $T$ = 300~K. We analyze the spectral components in the range [-100, 200] meV by a fit based on seven Gaussians (see Methods for details) and guided by the YIG's magnon dispersion along the [111] direction \cite{princep2017ful} (see Fig.~\ref{fig:1_1}\textbf{d}). We identify the elastic line at 0 meV (yellow), the acoustic magnon mode with parabolic dispersion at $\sim$  10 meV (blue), and the optical magnon modes grouped around $\sim$ 50 meV (light-blue) and $\sim$ 90 meV (green). The broad peak at $\sim$ 140 meV (grey) is ascribed to multi-magnon excitations\cite{Li2023prx}.  The spectral weight at \textcolor{black}{$\sim$ $-$10 meV and $\sim$ $-$50 meV}(red) indicates anti-Stokes annihilation of thermally excited magnons  \cite{nambu2020observepolarization}.

\section*{Spin current under temperature gradient}

Next, we present the RIXS data measured on a SSE device dedicated to producing a temperature gradient $ \bm \nabla T$ along the [111] direction, and schematized in Fig.~\ref{fig:2}\textbf{a}. Figure~\ref{fig:2}\textbf{b} displays the transverse voltage $V_{\text{SSE}}$ measured across the Pt electrode as a function of the temperature difference $ \Delta T$, induced by the thermal gradient $ \bm \nabla T$. The linear trend of $V_{\text{SSE}}$ confirms the presence of the spin Seebeck effect around $T \sim 300$~K \cite{chang2017role,uchida2012thermal,jaworski2010observation,kikkawa2015critical}.  The existence of a magnon spin current is deduced from the presence of $V_{\text{SSE}}$, that is 
caused by thermal spin pumping and inverse spin Hall effect  \cite{uchida2008observation, bauer2012spin}. 

Figure~\ref{fig:2}\textbf{c} presents the RIXS spectra measured on the device at 300~K under both equilibrium ($\Delta T_0$ = 0 K, purple dotted line) and non-equilibrium conditions ($\Delta T_1$ = 15 K, red dotted line), at $\mathbf{q}_{0.2}$.  At the presence of $\Delta T_1$, the RIXS spectrum mostly shows changes in intensity - e.g. no peak shift -- with respect to the one in equilibrium. In detail, we observe a broadening  of the $\sim50$~meV optical magnon mode and an increased spectral weight around the quasi-elastic line. 
The broadening in the $\sim50$~meV peak could be explained with an increase in the average temperature $T_{\text{av}} \sim T + \Delta T_1/2$, as spin waves tend to weaken and broaden at elevated temperatures \cite{lee2014asymmetry,RevModPhys_Dai_2015}. However, the increased spectral weight observed around the quasi-elastic region ([-10, 30] meV) cannot be easily explained by a temperature increase and is the first hint of a non-equilibrium effect. This is emphasized in Fig.~\ref{fig:2}\textbf{d} by the intensity difference $\Delta I_{300~\text{K}}^{15~\text{K}}$ (where $T$ = 300~K, and $\Delta T_1$ - $\Delta T_0$ = 15~K) obtained subtracting the spectrum at $\Delta T_0$ to the one at $\Delta T_1$.

To focus on this aspect, we performed similar non-equilibrium investigations at $T$ = 80~K, where the SSE of YIG is maximised \cite{guo2016influence,Concomitant_30K}. Figures ~\ref{fig:3 }\textbf{a-d} display the RIXS spectra under different temperature gradients: $\Delta T_2$ = 8.3~K, $\Delta T_3$ = 15.5~K, and $\Delta T_4$ = 24.8~K. We also compare results for scattering geometries corresponding to $\mathbf{q}_{0.2}$ = [0.2, 0.2, 0.2]~r.l.u. and 
$\mathbf{q}_{-0.2} = -\mathbf{q}_{0.2} $ = [-0.2, -0.2, -0.2]~r.l.u..
At $\mathbf{q}_{0.2}$, the RIXS intensity around $\sim$ 0~meV increases with $\Delta T$ (see Fig. \ref{fig:3 }\textbf{a}), while it decreases at $\mathbf{q}_{-0.2}$ (see Fig. \ref{fig:3 }\textbf{c}). A zoomed-in view on these spectral changes is presented in Figs.~\ref{fig:3 }\textbf{b} and~\ref{fig:3 }\textbf{d}, together with their differences $\Delta I$. Using the RIXS spectrum at $\Delta T_2$ = 8.3 K as a reference, we calculated $\Delta I^{\Delta T_i-\Delta T_2}_{80K}$, i.e. $\Delta I_{80~\text{K}}^{7.2~\text{K}}$ (light-blue dots) and $\Delta I_{80~\text{K}}^{16.5~\text{K}}$ (green dots).

The intensity differences at $\mathbf{q}_{0.2}$ and $\mathbf{q}_{-0.2}$ show well-defined peaks centered around $\sim$ 10~meV and a width corresponding to the instrumental energy resolution as emphasized by the  Gaussian fits, see Figs.~\ref{fig:3 }\textbf{c-d}. This result points to a single spectral component that changes under the effect of both $\Delta T$ and $\mathbf{q}$. We ascribe this to the acoustic magnon branch on the basis of its energy, and of its overall spectral enhancement at 80~K with respect to 300~K, 
see Fig.~\ref{fig:3 }\textbf{f}.

The integrals $S_{\mathbf{q}_{0.2}}$ and $S_{\mathbf{q}_{-0.2}}$ of each $\Delta I^{\Delta T_i-\Delta T_2}_{80K}$ computed over the energy region [-30,~40] meV, and plotted versus the temperature difference $\Delta T$ - $\Delta T_2$, summarize the main results (see Fig. ~\ref{fig:3 }\textbf{e}). The dashed guidelines highlight that both $S_{\mathbf{q}_{0.2}}$ and $S_{\mathbf{q}_{-0.2}}$ are linearly proportional to $\Delta T$ - $\Delta T_2$, but with opposite trends i.e.,  $S_{\mathbf{q}_{0.2}}$ increases while $S_{\mathbf{q}_{-0.2}}$ decreases with the thermal bias.

The linear dependence of the acoustic magnon peak intensity at 80 K on the applied temperature difference agrees with the behavior of $V_{\text{SSE}}$ (see Fig.~\ref{fig:2}\textbf{b}) measured on the same device, and cannot be explained by a global temperature increase (see Methods and Extended Data Fig. 2) . Similarly, it is not possible to explain the sign-asymmetry between the $S_{\mathbf{q}_{0.2}}$ and $S_{\mathbf{q}_{-0.2}}$ using equilibrium arguments \cite{ABaron}. We chose the momenta $\mathbf{q}_{0.2}$ and $\mathbf{q}_{-0.2}$ to be parallel or antiparallel to the direction of the magnon current density $\mathbf{J}_\text{S}$, respectively. The inversion of the sign of $S$ at $\mathbf{q}_{\pm 0.2}$ correlates with the direction of the magnon spin current. Moreover, the 
increase \textcolor{black}{of a factor of $\sim$ 3} in $\Delta I$ around the acoustic magnon region at 80~K with respect to 300~K for a comparable temperature difference (16.5~K at $T$ = 80~K versus 15~K at $T$= 300~K, see Fig.~\ref{fig:3 }\textbf{f}) agrees with the ratio of $V_{\text{SSE}}$ reported between the two temperatures \cite{guo2016influence,adachi2010gigantic,rezende2014magnon}. 
Finally, we exclude the effect of magnon accumulation and magnon depletion \cite{chemical_potential} at the edges of the sample as discussed in Extended Data Fig. 3.

\section*{Microscopic model of magnon spin current}
The presented results support the involvement of the acoustic magnon into the magnon spin current at $T$ = 80 K and 300 K. To validate our interpretation, we introduce in this section a microscopic picture of the magnon spin current in energy- and momentum-space, and we show that the acoustic magnon RIXS intensity as a function of $\mathbf{q}$ and $\Delta T$ directly probes the magnon distribution function \(f\), and thereby the magnon spin current. Magnons are Bosons, and at equilibrium obey the Planck distribution function,
\begin{equation}
\label{eqn:BE_model}
f_0(\omega_{\mathbf{q}}) = \frac{1}{e^{\hbar\omega_{\mathbf{q}}/(k_B T) }-1},
\end{equation}
where $k_B$ is the Boltzmann constant, and $f_0(\omega_\mathbf{q})=f_0(\omega_\mathbf{-q})$, see Fig.~\ref{fig:4}\textbf{a}.

The temperature gradient $\bm \nabla T$ forces the magnons to flow from the hot to the cold side, generating a magnon current. Assuming that $\mathbf{q}$ is parallel to the temperature gradient, the current corresponds to a shift of the entire magnon distribution along $\mathbf{q}$  such that $f(\mathbf{q}) > f(\mathbf{-q})$, as in Fig.~\ref{fig:4}\textbf{b}. The solution of the Boltzmann equation describes the non-equilibrium distribution function $f(\hbar\omega$,~$\mathbf{q})$, which governs the diffusive transport of spin and energy by magnons in magnetic insulators \cite{rezende2020fundamentals}. For small deviations from $f_0(\mathbf{q})$, the non-equilibrium distribution can be formulated using the linearized expression of the Boltzmann equation\cite{chemical_potential,Rezende_2018}:
\begin{equation}
    f(\mathbf{q}) - f_0(\mathbf{q}) = \frac{\tau_{\mathbf{q}}}{T} \frac{\hbar \omega_{\mathbf{q}}}{k_B T} 
    \frac{e^{\hbar\omega_{\mathbf{q}}/(k_B  T)}}{\left(e^{\hbar\omega_{\mathbf{q}}/(k_B  T)}-1\right)^2} \bm{\upnu_{\mathbf{q}}} \cdot \bm{\nabla} T,
    \label{eqn:fitting_model}
\end{equation}
where $\hbar\omega_{\textbf{q}}$ is the acoustic magnon energy, \textcolor{black}{$\tau_q$ is the magnon-conserving scattering, or simply, magnon relaxation time}, $\bm{\upnu_{\textbf{q}}}=\partial \omega_{\bf{q}}/\partial \bf{q}$ is the magnon group velocity that changes sign with \(\mathbf{q}\) and depend on the dispersion relation, and $\bm \nabla T$ is the temperature gradient.

 The calculated magnon distributions at 80~K for $\Delta T = 0$~K (solid line) and $\Delta T \neq 0$~K (dashed and dotted lines) are displayed in Fig.~\ref{fig:4}\textbf{c}, together with enlarged views close to the experimental momenta $\mathbf{q}_{\pm 0.2}$ (see Methods). The increase of the magnon occupation at $\mathbf{q}_{0.2}$ (parallel to $\mathbf{J}_\text{S}$) and decrease at $\mathbf{q}_{-0.2}$ (antiparallel to $\mathbf{J}_\text{S}$), as a function of $\Delta T$, qualitatively agrees with the experimental results 
 reported in Figs.~\ref{fig:3 }\textbf{c-d}. 
 This finding confirms the RIXS sensitivity to minute changes in the magnon distribution under non-equilibrium conditions, and unambiguously assigns the acoustic magnon mode as the main current carrier \cite{Boona_us}. Moreover, observing a thermally induced change of the acoustic magnon numbers at q$_{\pm 0.2}$ deep in the Brillouin zone complements previously reported low-temperature studies limited to $q\sim 0$ \cite{Boona_us,Long_lifetime}. 

By quantitatively comparing our experimental results and calculations, we can extract the momentum-dependent magnon relaxation time $\tau_{\bf{q}}$ through a single-parameter fit of the data. 
The squares (diamonds) in Fig.~\ref{fig:4}\textbf{d} are the energy-integrated RIXS intensities  $A_{{\bf{q}}_{0.2}}$ ($A_{{\bf{q}}_{-0.2}}$)  of \textcolor{black}{Gaussian fits of the} acoustic magnons at ${\bf q}_{0.2}$ (${\bf q}_{-0.2}$) displayed as a function of $\Delta T$ \textcolor{black}{(see Extended Data Fig. 4)}. 
Here $A_{\bf{q}}$ is proportional to $f({\bf{q}})$ (see  Extended Data Fig. 4 and Eq.~\ref{eqn:crosssection}), and  $\tau_{\bf q}= \tau_{\bf -q}$. \textcolor{black}{The difference between $A_{\bf{q}}$ values for different $\Delta T$s refines the values of the raw $S_q$ integrals.} 
The solid lines in Fig.~\ref{fig:4}\textbf{d} report the best fit of the data based on Eq.~\ref{eqn:fitting_model} using a magnon relaxation time of $\tau_{q}$ = 58 $\pm$ 4 ns at $|\bf{q_{\pm 0.2}}|$. \textcolor{black}{The error bars are given by the least squares method.} This number is compared with estimates from previous studies as summarized  in the Extended Data Table 1.  
\textcolor{black}{The extracted value $\tau_{\textbf{q}_{0.2}}\sim $58 ns is shorter than the only other report of a direct measurement close to the Brillouin zone center \cite{Long_lifetime}, i.e. $\tau_{\mathbf{q}_{0}} \sim 2-60 \, \mathrm{\mu s}$, consistent with the evidence that the magnon scattering rate increases with energy and momentum \cite{largeq480ns, kikkawa2015critical}}, see Methods for more details.


\section*{Perspective}

The direct spectroscopic \textcolor{black}{detection} 
--  energy and momentum resolved -- of the magnon current by RIXS should stimulate the theory and \textcolor{black}{the tomography} 
of material- and momentum-dependent relaxation times \textcolor{black}{in reciprocal space}. Additionally, this method holds promise to open many research avenues. %
RIXS 
could guide the optimization of magnonic devices by film thickness \cite{pelliciari2021tuning} and choice of materials. RIXS may help to understand the giant magnon spin conductivity of ultrathin YIG films close to the two-dimensional limit \cite{wei2022giantNM}, encouraging the study of two-dimensional magnons in van der Waals mono- and multilayers. Another promising playground for RIXS should be the excitations of multiferroic materials such as electromagnons \cite{kubacka2014}.  
More generally, our RIXS experiments can be extended to other types of exotic chargeless transport \cite{gish2024van, han2020spin, qian2021phonon, franchini2021polarons,choi2023nature}, by unveiling the currents of the underlying carriers (magnons, phonons, excitons, orbitons, polarons) and their decay processes.


\begin{figure*}[!h]
\centering
\includegraphics[width=\linewidth]{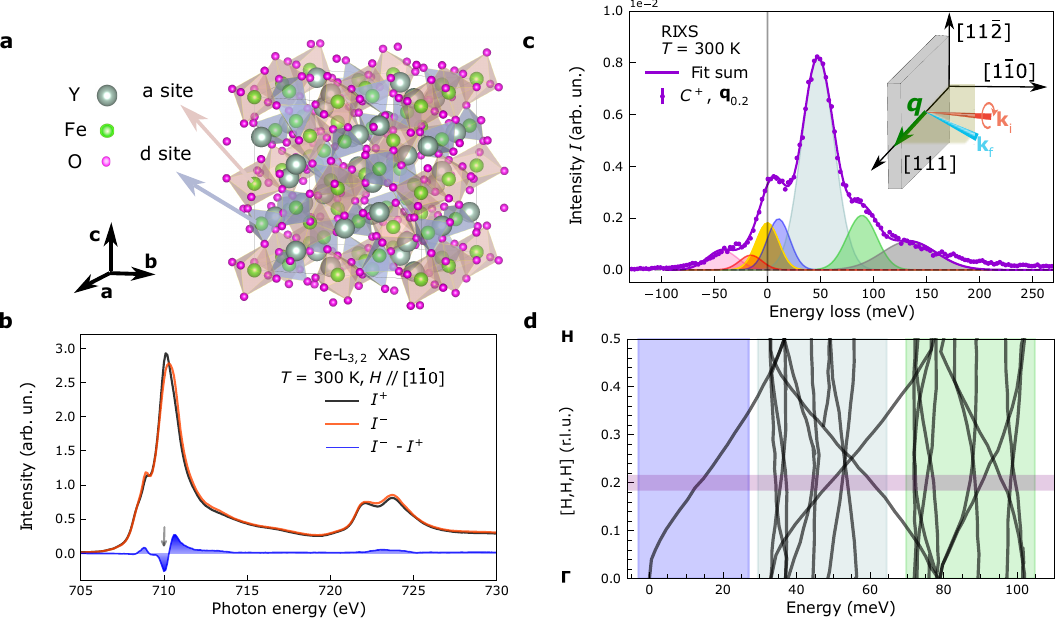}
\caption{\textbf{Structural, spectroscopic, and magnetic properties of YIG at 300~K.} \textbf{a,} YIG unit cell crystal structure. Tetrahedral, majority spin Fe sites are represented in light blue; octahedral, minority  spin Fe sites in light pink. \textbf{b,} Fe L$_{3,2}$-edges XAS spectra $I^+$ (black line) and $I^-$ (red line) measured respectively for $C^+$ and $C^-$ beam polarization, in total fluorescence yield (TFY) mode. A magnetic field $H$ = 700 Oe is applied along the [1$\bar{1}$0] direction, anti-parallel to the incoming beam. The XMCD signal  $I^- - I^+$ is displayed in blue. The grey arrow identifies the tetrahedral (majority) Fe site resonance (710 eV), used in all RIXS measurements. \textbf{c,} RIXS spectrum (purple dots) measured at ${\bf q}_{0.2}$ = [0.2, 0.2, 0.2] r.l.u. with an incident energy of 710~eV, and $C^+$ polarization. The purple solid line is the sum of the seven fitted Gaussians (color-shaded areas). The inset sketches the scattering geometry with incoming ($\mathbf{k}_{\text i}$) and scattered ($\mathbf{k}_{\text f}$) wave vectors and momentum transfer $\mathbf{q}=\mathbf{k}_{\text f}-\mathbf{k}_{\text i}$. \textbf{d,} Calculated magnon dispersion of YIG, following Ref. \cite{princep2017ful}. The horizontal shaded area highlights the momentum ${\bf q}_{0.2}$ = [0.2, 0.2, 0.2] r.l.u.. The vertical shaded areas highlight the assignment of the peaks observed in the RIXS spectrum at panel \textbf{c}, using the same color code: blue for acoustic and light-blue and green for optical magnons. The width of the colored shades agrees with that of the Gaussians used to fit the RIXS spectrum.}
\label{fig:1_1}
\end{figure*}

\begin{figure*}[!h]
\centering
\includegraphics[width=\linewidth]{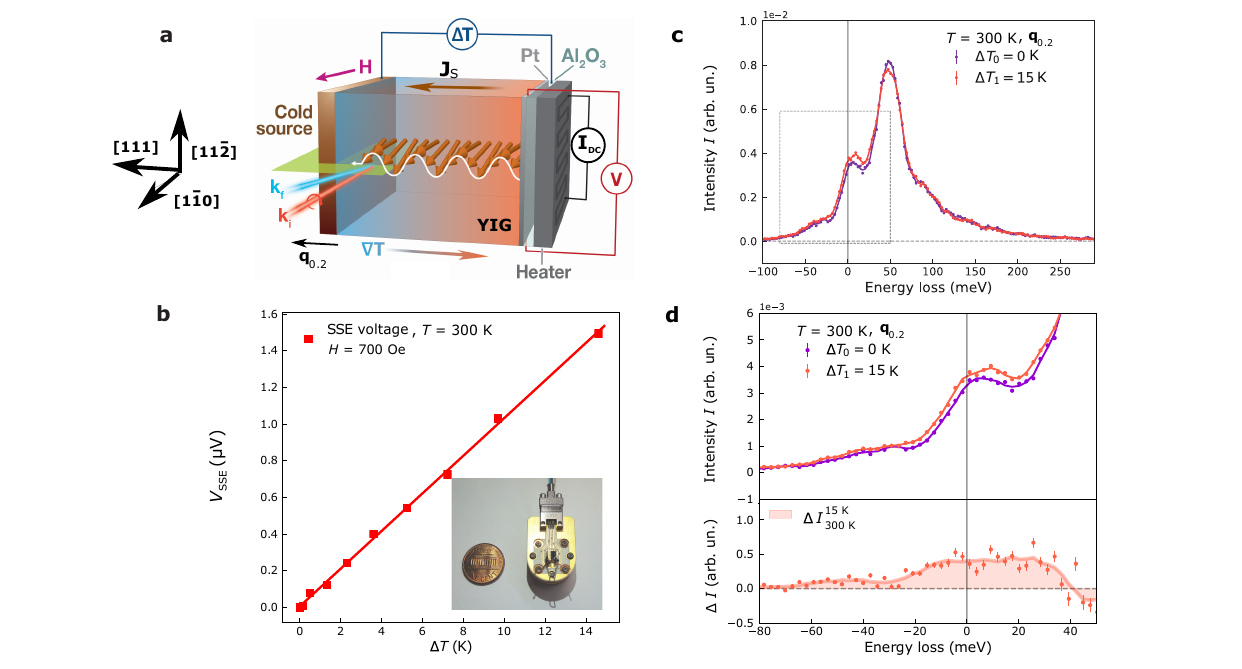}
\caption {\textbf{Spin Seebeck device, SSE voltage and RIXS spectra at 300 K.}  \textbf{a,} Schematic of the SSE device used in the RIXS experiment and based on a YIG single crystal (credit to Tiffany Bowman, Brookhaven National Laboratory). The cold source sets the global temperature $T$, while the heater provides the temperature gradient $\bm \nabla T$ along the [$\bar{1}$$\bar{1}$$\bar{1}$] direction. A thermocouple monitors the temperature difference $\Delta T$ across the sample. \textcolor{black}{The corresponding temperature gradients $\nabla T = \Delta T / L$ follow from the sample length $L=1.3$ mm.}
The ISHE in Pt thin film generates a transverse spin Seebeck voltage $V_{\text {SSE}}$. A  magnetic field \(H=700\) Oe  saturates the magnetization along the [1$\bar{1}$0] direction. The central red arrows represent the spin waves carrying a magnon spin current. The RIXS scattering plane (green color) is defined by the incoming beam \(\bf k_{\rm{i}}\)  (orange beam), and the outgoing beam \(\bf k_{\rm{f}}\) (blue beam) at $\bf{q}_{0.2}=\bf{k}_{\rm{f}}-\bf k_{\rm{i}}$. The RIXS measurements were performed at the cold end of the temperature gradient. \textbf{b,} $V_{\text {SSE}}$ measured as a function of $\Delta T$, at 300 K. A picture of the device is included as an inset. \textbf{c,} RIXS spectra at momentum transfer ${\bf q}_{0.2}$ = [0.2, 0.2,0.2] r.l.u. at equilibrium $\Delta T_0$ = 0 K (purple dots) and non-equilibrium $\Delta T_1$ = 15 K (red dots). \textbf{d,} Enlarged view of the dashed-area in Fig. 2\textbf{c}, including the intensity difference $\Delta I_{300~\text{K}}^{15~\text{K}}$ (dots). The solid line is a 3-point smoothing of $\Delta I_{300~\text{K}}^{15~\text{K}}$, and the enclosed area is filled in light red.}
\label{fig:2}
\end{figure*}

\begin{figure*}[!h]
\centering
\includegraphics[width=\linewidth]{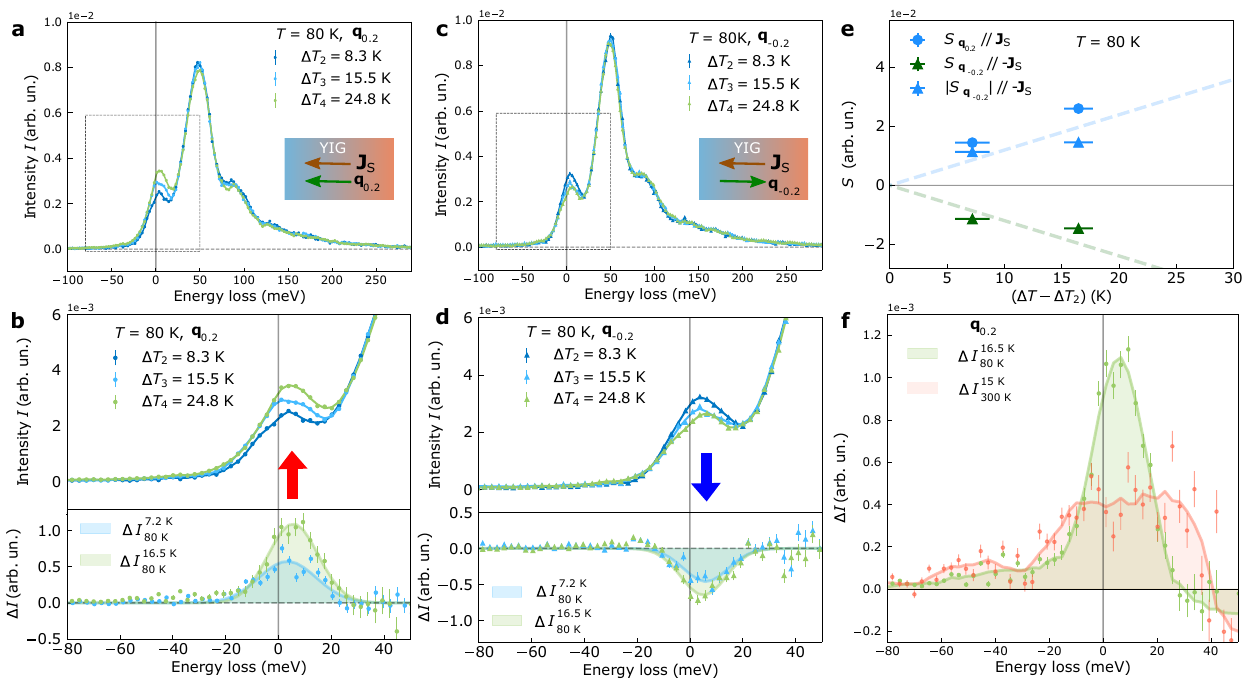}
\caption{\textbf{Non-equilibrium RIXS measurement at 80 K.} \textbf{a, c} RIXS spectra recorded for momentum ${\bf q}_{0.2}$ = [0.2, 0.2, 0.2] r.l.u.,  and ${\bf q}_{ -0.2}$ = [-0.2, -0.2, -0.2] r.l.u., at $T$ = 80 K and at three temperature differences $\Delta T_2$, $\Delta T_3$, and $\Delta T_4$. \textcolor{black}{Our device set-up (see Methods), does not allow measuring spectra for $\Delta T$ = 0 K at $T$ = 80 K}. \textbf{b, d} Enlarged  views of the dashed-area in Fig. 3\textbf{a} and Fig. 3\textbf{c}, respectively. The dots in the lower panels are the intensity differences $\Delta I_{80~\text{K}}^{7.2~\text{K}}$ ($\Delta I_{80~\text{K}}^{16.5~\text{K}}$) at  temperature difference $\Delta T_3$ -$\Delta T_2$ = 7.2 K ($\Delta T_4$ -$\Delta T_2$ = 16.5 K). These data are fit by a single Gaussian (solid lines) with a width determined by the experimental resolution. \textbf{e,} 
\textcolor{black}{$S$, integral of the spectral differences} $\Delta I_{80~\text{K}}^{7.2~\text{K}}$ and $\Delta I_{80~\text{K}}^{16.5~\text{K}}$ at 80 K in the window [-30, 40] meV versus $\Delta T -\Delta T_2 $, for ${\bf q}_{0.2}$ (circles) and ${\bf q}_{-0.2}$ (triangles). The dashed trend line are a linear guide for the eye. \textbf{f,} Comparison of $\Delta I_{80~\text{K}}^{16.5~\text{K}}$ (green dots) and $\Delta I_{300~\text{K}}^{15~\text{K}}$ (pink dots) at ${\bf q}_{0.2}$.} 
\label{fig:3 }
\end{figure*}

\begin{figure*}[!h]
\centering
\includegraphics[width=\linewidth]{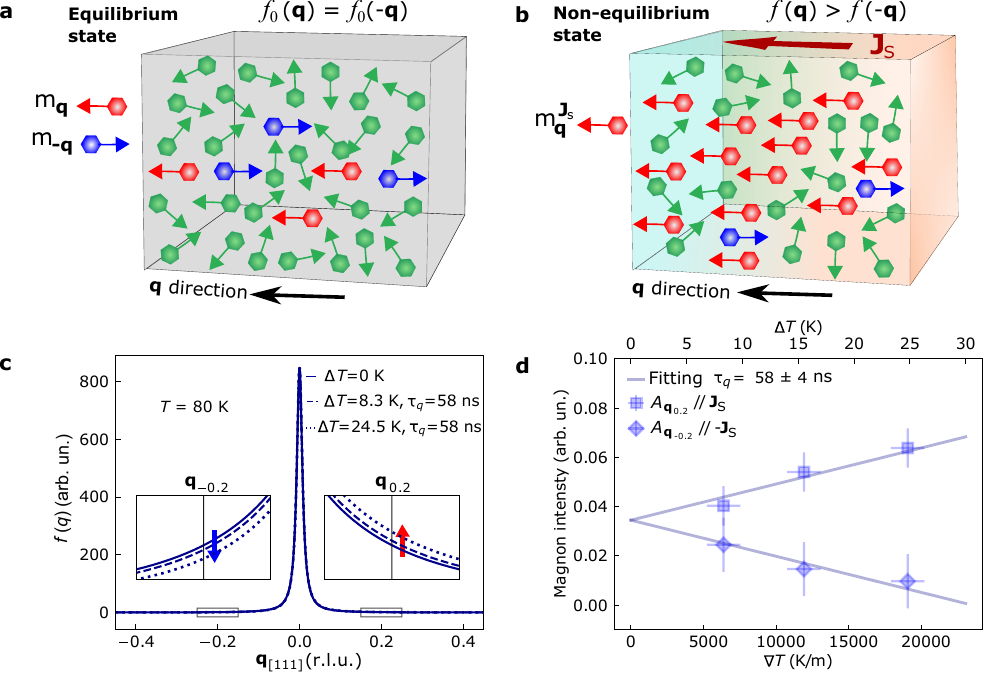}
\caption{\textbf{ Magnon distributions function.}  \textbf{a,} Real-space sketch of the magnon gas at equilibrium, and  \textbf{b,} under a temperature gradient. The arrowed polygons represent magnon wave packets. In equilibrium (green polygons), their distribution is symmetric in reciprocal space {\bf q}, $f_0({\bf q})$ = $f_0({\bf -q})$, so the same number of magnons moves to the left as to the right. A perturbation in the \(\mathbf{q}\) direction that breaks inversion symmetry shifts the magnon distribution such that $f({\bf q})>f({\bf -q})$. This leads to a magnon current {\bf J}$_\text{S}$ (red arrowed-polygons) parallel to {\bf q}. \textbf{c,} The Boltzmann distribution function for YIG acoustic magnons at 80 K in equilibrium and non-equilibrium with a constant relaxation time \textcolor{black}{of 58 ns}. \textbf{d,} Fit of the RIXS magnon intensity $A_{{\bf{q}}_{0.2}}$ and $A_{{\bf{q}}_{-0.2}}$ as a function of $\Delta T$ by the solution of the linearized Boltzmann equation. 
The error bars indicate the experimental standard deviations.}
\label{fig:4}
\end{figure*}

\newpage
\clearpage

\section*{Methods}

\subsection*{Crystal details}
Yttrium Iron Garnet is a ferrimagnetic insulator with $T_\text{C}$ $\sim$ 560 K 
with a large, cubic unit cell ($a$ = 12.3755~\AA) of 40 Fe atoms \cite{Joe2016Thermal,Joe2019Semiquantum,princep2017ful}. The reciprocal lattice unit for YIG corresponds to 1 r.l.u. =  \(2 \sqrt{3}\pi / 1.24\) nm\(^{-1}\).
The magnetic Fe$^{3+}$ ions ([Ar]3$d^5$) are in the high-spin electronic configuration S = 5/2 and occupy the octrahedral ($a$) and tetrahedral ($d$) sites in the oxygen scaffold with a ratio of 2 : 3. The spins of the Fe local moments of majority $d$-sites and minority $a$-sites are antiparallel with a net magnetization  along the [111] easy axis below $T_\text{C}$. 

Our sample is a single crystal slab with [1$\bar{1}$0] surface normal and [111] axis in the film and scattering plane purchased from MTI Corporation.  

\subsection*{X-ray Absorption Spectroscopy}

\textcolor{black}{The Fe L$_{3,2}$-edge x-ray absorption spectroscopy (XAS) in the XMCD mode is highly sensitive to the spin orientation and chemical environment  \cite{Vasili2017XMCD}. 
Figure~\ref{fig:1_1}\textbf{b} shows the room temperature XAS $I^+$ (black line) and $I^-$ (red line), respectively 
measured with positive $(C^+)$ and negative $(C^-)$ circular light polarization. The XMCD signal ($I^--I^+$) (blue line) agrees with previous measurements \cite{Vasili2017XMCD}. The negative XMCD response at $\sim710$~eV comes from the majority tetrahedral $d$-sites and corresponds to the transition from filled Fe 2$p_{3/2}$ core levels to unoccupied Fe 3$d_{e}$ levels, split by the crystal field from the 3$d_{t_{2g}}$ ones. We used this resonant energy for the RIXS measurements, since it maximizes the sensitivity of the RIXS spectrum to magnetic effects \cite{elnaggar2019magnetic}.}

\subsection*{Fitting the RIXS spectrum at 300K}
We fitted the room-temperature RIXS spectrum of YIG at ${\bf q}_{0.2}$ = [0.2, 0.2, 0.2]~r.l.u. by the energy position and amplitude of 7 Gaussians. Only the widths of the acoustic magnon modes (red and blue areas) and the elastic peak (gold area) are fixed at \(w\) (see Fig. 1\textbf{c}), reflecting the instrumental energy resolution, whereas the linewidths of the optical and multi-magnon modes are adjusted to the data. The resulting widths are illustrated by the shaded areas overlaying the acoustic (blue) and optical (light-blue and green) magnon bands in Fig. 1\textbf{d}.

\subsection*{Device fabrication}
The device in this RIXS study is a YIG single crystal, contained between a cold source -- cryostat extension that stabilizes the temperature $T$ -- and a hot source -- heater, at a temperature $T + \Delta T$ activated by the charge current I$_\text{DC}$ (see Fig. ~\ref{fig:2}\textbf{a}). We cut the YIG single crystal into a quasi-cuboid with size 1.3 $\times$ 1.38  $\times$ 2.46 mm with a diamond wire saw. The (111) and
 (1$\bar{1}$0) sample faces were polished with MultiPrep System down to a surface roughness of 50 nm: the (111) face was covered with 15 nm of Pt thin film deposited using a E-beam evaporator for the SSE voltage measurement, while the (1$\bar{1}$0) face was used for the RIXS measurement. A magnetic field of 700 Oe was applied along the [1$\bar{1}$0] direction. A thin film resistor heater (purchased from KOA Speer Company) was glued on the YIG single crystal with a sapphire spacer that ensured electrical isolation and good thermal contact. A type-E chromel/constantan thermocouple from LakeShore Cryotronics between the heater and the cold source monitored $\Delta T$. Due to the configuration of the SSE device, we could not achieve the equilibrium state ($\Delta T$ = 0 K) at $T$ = 80 K as the hot-end of the crystal was not in thermal contact with the cryostat. 

\subsection*{Electrical measurements}
The SSE voltage $V_{\text{SSE}}$ was measured with a Keithely 2182A nanovoltmeter. The heater was activated by a charge current $\rm{I_{DC}}$ ( $<$ 50 mA) sourced by a Keithely 2636B in the current source mode for channel A, while the temperature gradient was extracted from the thermocouple voltage differential measured by the Keithely 2636B in the voltage measure mode for channel B. The voltage was then converted to temperature difference with the known Seebeck coefficient of the type-E thermocouple. The extracted $\Delta T$ scaled like $\rm{I_{DC}}^2$ and the $V_{\text{SSE}}$ was linear with $\Delta T$ (see Fig. ~\ref{fig:2}\textbf{b}).  During the RIXS experiment, both $\Delta T$ and $V_{\text{SSE}}$ were constantly monitored to ensure proper and continuous operation of the SSE device over several hours.

\subsection*{XAS and RIXS measurements}
XAS was measured at the SIX 2-ID beamline of National Synchrotron Light Source II (NSLS-II) with a resolution of $\sim$50 meV, in TFY mode using a photo-diode. Positive and negative circularly polarized light of the incoming beam was used to acquire the XMCD signal, at the presence of 700 Oe magnetic field, at 300~K.

The RIXS measurements were performed at SIX 2-ID beamline of the NSLS-II using the high-resolution Centurion RIXS spectrometer \cite{Joseph2016}. The scattering plane was [1$\bar{1}$0] $\times$ [111] in reciprocal space.   The resonant energy was tuned to Fe-L$_3$ edge with energy resolution around 25 meV, defined as the full-width at half-maximum of the non-resonant diffuse scattering from the silver paint. RIXS data were collected at scattering angle 150$ ^{\circ}$ with momentum transfer $|{\textbf Q}|$ = 0.695  \AA $^{-1}$. The incident angle was set to $\theta_{\text{in}}$ = 90$^{\circ}$ and $\theta_{\text{in}}$ = 60$^{\circ}$, respectively enabling a momentum transfer of ${\bf q}_{0.2}$ = [0.2, 0.2, 0.2]~r.l.u. and ${\bf q}_{-0.2}$ = [-0.2, -0.2, -0.2]~r.l.u. along the [111] direction. 
The RIXS incident photon energy was set to the 2$p_{3/2}$ $\rightarrow$ 3$d_{e}$ resonance of the majority Fe $d$-sites with $C^+$ polarized light, to enhance their magnetic response \cite{elnaggar2019magnetic}. 
The elastic peaks were aligned after Gaussian fitting and all the intensities were normalized by the total spectral weight.

 \subsection*{Parameters for magnon distribution}
Figure~\ref{fig:4}\textbf{c} shows  Eq.~\ref{eqn:fitting_model} calculated for $T=80$~K, where the temperature gradient ${\bm \nabla} T$ is obtained considering the measured $\Delta T$ values and the sample width along the [111] direction. We considered the single acoustic magnon mode of YIG ($E\leq~30$~meV) with parabolic dispersion $E(q)= Dq^2$, where \(D = 7.42\times 10^{-40}\mathrm{Jm}^2\) is the spin wave stiffness, since non-parabolicities are weak and the optical modes are hardly occupied at $T\leq~300$~K. As a starting value, we set $\tau_{q}$ = 58 ns in Fig.~\ref{fig:4}{\bf c}.

\subsection*{RIXS experimental geometry}

RIXS probes the energy $\hbar \omega$  and the momentum $\hbar \mathbf{k}$ of elementary excitations, i.e. magnons in this study, by scattering photons with $\mathbf{k} =  \mathbf{k}_\mathrm{f} - \mathbf{k}_\text{i}$ and $\omega =\omega_\text{i} - \omega_\text{f}$, where $\hbar \mathbf{k}_\text{i}$ ($\hbar \mathbf{k}_\text{f}$) and $\hbar \omega_\text{i}$ ($\hbar \omega_\text{f}$) are the momentum and energy of the incoming (outgoing) photons, respectively.
We collected RIXS spectra for two different scattering geometries: $\theta_{\text{in}}=90^{\circ}$/$2 \Theta=150^{\circ}$ (see Extended Data Fig. \ref{fig:momentum}\textbf{a}) and $\theta_{\text{in}}=60^{\circ}$/$2 \Theta=150^{\circ}$ (see Extended Data Fig. \ref{fig:momentum}\textbf{b}), with the scattering plane defined by [H,H,H] $\times$ [H,$\Bar{\text {H}}$,0].
Due to YIG's large unit cell ($a$ = $b$ = $c$ = 12.3755 \AA) soft x-rays reach deeply into the Brillouin zone. With $\mathbf{Q} = 2 \mathbf{k}_\text{i} \sin(2 \Theta /2)$ and scattering angle $2 \Theta=150^{\circ}$, the momentum transfer $|\mathbf{Q}_1|=|\mathbf{Q}_2|$ = 0.695 \AA $^{-1}$ is close to the (1$\bar{1}$0) Bragg peak at $|\mathbf{G}_{1 \bar{1} 0}|=0.718$ \AA $^{-1}$. For this reason, we can formulate the momentum transfer $\mathbf{Q}_1 = \mathbf{G}_{1 \bar{1} 0} + \mathbf{q} = \mathbf{q}$ and $\mathbf{Q}_2 = \mathbf{G}_{1 \bar{1} 0} + (-\mathbf{q}) = -\mathbf{q}$.  In the two scattering geometries, the angle between the momentum transfer  $\mathbf{Q}_1$ ($\mathbf{Q}_2$) and $\mathbf{G}_{1\bar{1}0}$ is  15$^\circ$, so $\mathbf{q}$ is nearly parallel to the [111] direction. The inelastic momentum transfer used in the main text is \(\pm \mathbf{q}\) as measured from the Bragg peak $\mathbf{G}_{1 \bar{1} 0}$ in the direction parallel to the [111] axis, as shown by the green arrows in Extended Data Fig. \ref{fig:momentum}. For  $\theta_{\text{in}}=90^{\circ} (60 ^{\circ})$ therefore $\mathbf{q}$~=~$\mathbf{q}_{0.2}$ = [0.2,0.2,0.2]  ($-\mathbf{q}$~=~$\mathbf{q}_{-0.2}$ = [-0.2,-0.2,-0.2]) r.l.u.

\subsection*{Temperature dependence of RIXS spectra}
Temperature affects the elementary excitation spectra \cite{Ament_rixs_review,RevModPhys_Dai_2015} in various ways. For example, increasing sample temperature 
suppresses the photon scattering by magnons\cite{lee2014asymmetry,manley2019intrinsic,boschini2021dynamic,nambu2021neutron,Coulomb}. Since we heat the sample on one side and cool it on the other, we must assess the effect of the average temperature increase  ($T_{\text{av}}$). To this end, we measured RIXS on a reference sample directly attached to the cold finger of the cryostat at two different temperatures but without temperature gradients. 
We measured RIXS at $T = 80$ K (same temperature selected for the spin Seebeck device) and  $T = 107$ K: the temperature difference of 27 K is close to the largest temperature bias over the device, $\Delta T_4$ = 24.8 K at 80 K. Both were cut from the same piece of single crystal, with similar shape and orientation, and RIXS scattering geometry was the same, see Fig. 1\textbf{c} in the main text or Extended Data Fig. \ref{fig:momentum}\textbf{a}.

Extended Data Fig. \ref{fig:ref}\textbf{a} displays the low-energy portion of the RIXS spectrum at 80 K (107 K) in the light blue (dark blue) scattered line, while their difference is shown in Extended Data Fig. \ref{fig:ref}\textbf{b}. The measured RIXS intensities do not depend significantly  on the temperature differences  $\Delta T \leq 27~$K. An increase in the average temperature can therefore not cause the RIXS results reported  in Fig. 3 of the main text.

\subsection*{Spin current at different positions}

The spin Seebeck effect generates a magnon accumulation (depletion) at the cold (hot) interface \cite{chemical_potential,Differences,AIP_nonlocal} on the scale of the magnon relaxation length $\lambda$\cite{Evidence,Intrinsic,kehlberger2015length,Controlling}. The deviation of the local magnon density from the equilibrium state $\rho$ is an accumulation for $\rho > 0$  and a depletion for $\rho < 0$.
It is governed by the one-dimensional diffusion equation 
\begin{equation}
\frac{\partial ^2 \rho}{\partial x^2} = \frac{\rho}{\lambda^2}.
\end{equation}
Under a uniform constant temperature 
gradient $\nabla$T along \textit{x} \cite{chemical_potential,nonlocal_diffusion_PRB2016,zhang_2012PRL,zhang_2012PRB} its general solution reads $\rho (x)= Ae^{-x/\lambda}+Be^{x/\lambda}$, where A and B are constants that depend on the boundary conditions.

In Extended Data Fig. ~\ref{fig:pos}{\bf a} we sketch of our YIG-based device with a total length $L$ = 1300 $\mu$m. Extended Data Fig. ~\ref{fig:pos}{\bf b} shows the magnon accumulation/depletion $\rho(x)$ calculated for our device for magnon relaxation lengths $\lambda= 2 ~\mu $m (red solid line) and $\lambda= 10 ~\mu$m (black solid line). Since $L\gg $ $\lambda$, we find that $\rho(x)$ = 0 in most of the bulk in which the spin current is generated solely by the temperature gradient. In the following, we argue that a magnon diffusion current is indeed the same over the entire sample except perhaps very close to the edges.

All RIXS measurements in the main text were collected 160 $\mu$m away from the cold edge of the sample, the so called ``cold point" in Extended Data Fig. ~\ref{fig:pos}{\bf a}. Given that this distance exceeds the magnon relaxation length in YIG \cite{Evidence,Intrinsic,kehlberger2015length,Controlling} (usually $\lambda \leq 10\mu m$), it is reasonable to assume that the variation observed in the RIXS intensity versus $\Delta T$ is independent from the magnon accumulation effect. We confirm this expectation by measuring RIXS at a  ``hot point'' on the sample  (160 $\mu$m from the hot interface) under the same conditions as before (Extended Data Fig. ~\ref{fig:pos}{\bf a}). 

In Extended Data Fig.  \ref{fig:pos}{\bf c} ({\bf e}), the dark and light blue dots represent spectra collected for temperature differences $\Delta T_2$ and $\Delta T_4$ close to the hot point (cold point). The intensity difference at the hot point (cold point) between spectra at $\Delta T_2$ and $\Delta T_4$, as seen in Extended Data Fig.  \ref{fig:pos}{\bf d} ({\bf f}), is a resolution-limited Gaussian peak (solid line) indicating the absence of a spin current driven by a magnon accumulation/depletion rather than a temperature gradient.

\subsection*{Magnon distributions function}

We extract the momentum \textbf{q}-resolved magnon relaxation time $\tau_{\mathbf{q}}$ from the experimental spectra via the magnon distribution function  $f(\mathbf{q})$ under an applied temperature gradient, calculated by the linearized Boltzmann equation for magnons. When \(|f-f_0| \ll 1 \), where $f_0(\mathbf{q})$ is the thermal equilibrium distribution,
\cite{chemical_potential}

\begin{equation}
    f(\mathbf{q}) - f_0(\mathbf{q}) \approx \tau_{\mathbf{q}} \frac{\partial}{\partial \hbar \omega_{\mathbf{q}}}\left(\frac{1}{e^{\hbar\omega_{\mathbf{q}}/(k_B T_p)}-1}\right) \bm{\upnu_{\bf{q}}}  \cdot \left( {\bm \nabla} \mu_m + \hbar \omega_{\mathbf{q}} \frac{{\bm \nabla} T_m}{T_p} \right).
    \label{BZlin}
\end{equation}

Here $\hbar\omega_{\mathbf{q}}$ is the magnon dispersion, $\tau_{\mathbf{q}}$ is the relaxation time, $\bm{\upnu}_{\mathbf{q}} = \partial \omega _\mathbf{q} / \partial \mathbf{q}$ is the magnon group velocity, $\bm \nabla $ is a spatial gradient, $\mu_m$ is the magnon chemical potential, $T_m$ is the magnon temperature and $T_p$ is the phonon temperature.

We compare the model with RIXS measurements on a spot separated $\sim$160 $\mu$m from the cold end of the sample at which any magnon accumulation $\mu_m=0$, see Extended Data Fig. 3. Since at high temperature the magnon and phonon scattering is strong, they are at local equilibrium with $T_m = T_p = T$. Equation \ref{BZlin} then simplifies to: 

\begin{equation}
    f(\mathbf{q}) - f_0(\mathbf{q}) = \frac{\tau_{\mathbf{q}}}{T} \frac{\hbar \omega_{\mathbf{q}}}{k_B T} 
    \frac{e^{\hbar\omega_{\mathbf{q}}/(k_B T)}}{\left(e^{\hbar\omega_{\mathbf{q}}/(k_B T)}-1\right)^2} \bm{\upnu_{\bf{q}}} \cdot \bm{\nabla} T,
    \label{mag_dis}
\end{equation}

While magnon dispersion in YIG is complex (see Fig. 1\textbf{d} in the main text), only the  magnons up to THz are mobile and contribute to transport, viz. those in the lowest acoustic branch with parabolic and isotropic dispersion:
\begin{equation}
    \varepsilon(q) = \hbar\omega(q) = \varepsilon_0 + Dq^2
    \label{energy}
\end{equation}
where $\varepsilon_0 = g \mu_B B_z$ is the energy gap due to an applied field with $B_z = 700$~Oe, $g=2.002$ represents the electron g-factor and $\mu_B = 9.274\times10^{-24}$~J~T$^{-1}$ is the Bohr magneton. We adopt the spin wave stiffness $D=7.42\times 10^{-40}\mathrm{Jm}^2$ ~\cite{rezende2015thermal} and $q = |\textbf{q}|$. \textcolor{black}{This is valid when the ferromagnetic dispersion is parabolic, and Eq. \ref{energy} is a reasonable assumption for small \textbf{q} values.} 
Using Eq. \ref{energy}, we can formulate the group velocity as:

\begin{equation}
  \bm{\nabla} T = \hat{\mathbf{x}} \frac{\partial T }{\partial x }\rightarrow \frac{\partial \omega_{\mathbf{q}}}{\partial\mathbf{q}}  \cdot \bm{\nabla} T=\frac{1}{\hbar}\frac{\partial \varepsilon(q)}{\partial q} \frac{\partial T }{\partial x }\frac{\textbf{q}}{q}\cdot \hat{\mathbf{x}} = \frac{2Dq \cos \theta}{\hbar} \frac{\partial T}{\partial x}=\frac{2Dq_x}{\hbar} \frac{\partial T}{\partial x}, 
  \end{equation}

Equation (2) in the main text with `+' (`-') for  \textbf{q} parallel (antiparallel) to the temperature gradient then reads
\begin{equation}\label{fq_dis}
    f(\textbf{q}) - f_0(q) = \frac{2Dq_x\tau_q}{\hbar} \frac{(\varepsilon_0 + Dq^2)}{k_B T^2} 
    \frac{e^{(\varepsilon_0 + Dq^2)/(k_B T)}}{\left(e^{(\varepsilon_0 + Dq^2)/(k_B T)}-1\right)^2} \frac{\partial T }{\partial x }  ~ \
\end{equation}

\subsection*{Fitting of the RIXS data}

We fitted the RIXS spectra at $T$ = 80 K in the non-equilibrium state with five Gaussian peaks (elastic, acoustic magnon, two optical magnon branches, multimagnon), similar to the procedure in the main text at $T$ = 300 K (see Fig. 1 $\bf{c}$), but without the anti-Stokes contribution.  The lower temperature $T$ = 80 K freezes out the magnons and suppresses the anti-Stokes channel. We fix the width of the elastic peak to the experimental resolution, $\Delta E$ = 25 meV and assume that its amplitude does not depend on the temperature gradient as established by Extended Data Fig. ~\ref{fig:ref}.  We allow all other parameters to vary freely. 

Extended Data Fig.~\ref{fig:fitting}\textbf{a} shows RIXS spectra measured at ${\bf q}_{0.2}$ and three temperature differences ($\Delta T_2$, $\Delta T_3$, and $\Delta T_4$). The triangles represent the experimental data, the solid lines depict the fit, and the shaded areas mark the individual  contributions. Extended Data Fig. \ref{fig:fitting}\textbf{b} shows the results for ${\bf q}_{-0.2}$. We conclude that the intensity ascribed to the acoustic magnon branch increases with temperature difference for ${\bf q}_{0.2}$, but decreases for ${\bf q}_{-0.2}$. %
\textcolor{black}{We attribute t}he difference in the elastic intensity between the two $\bf{q}$-values 
to phenomena unrelated to the magnon spin current, such as \textcolor{black}{elastic scattering} at surface roughness that is enhanced by the \textcolor{black}{larger} projected beam at $\theta_{in}=60^\circ$ \textcolor{black}{(${\bf q}_{-0.2}$), relative to $\theta_{in}=90^\circ$ (${\bf q}_{0.2}$)}.

The integral of the Gaussian that fits the acoustic magnon branch intensity is $A_{\bf{q}}$, plotted as a function of temperature difference in the main text Fig. 4\textbf{d}. 
The RIXS scattering intensity is proportional to the magnon distribution function, through a geometrical factor and an absolute conversion factor\cite{Bisogni2014prl,jia2014persistent,Groot1998density,Robarts2021spin_susceptibility}. We assume that the geometrical factor does not significantly change between the two  configurations ${\bf q}_{0.2}$and ${\bf q}_{-0.2}$ with scattering angles  $\theta=60^\circ$ and $\theta=90^\circ$ (see Extended Data Fig. ~\ref{fig:momentum}), respectively. Therefore $A_{\bf{q}}=I_0 f(\bf{q})$, where \(I_0\) is the RIXS cross section and $f({\bf q})$ is the non-equilibrium magnon distribution ( Eq. \ref{fq_dis}).  We therefore model the $A_{\bf{q}}$ displayed in the main text Fig. 4\textbf{d} as
\begin{equation}
   A_{{q}} = I_0 \left[ \frac{2Dq_x\tau_q}{\hbar} \frac{(\varepsilon_0 + Dq^2)}{k_B T^2} 
    \frac{e^{(\varepsilon_0 + Dq^2)/(k_B T)}}{\left(e^{(\varepsilon_0 + Dq^2)/(k_B T)}-1\right)^2} \frac{\partial T}{\partial x} + f_0({q})  \right]
    \label{eq:fitting}
\end{equation}
Here $q$ = $|{\bf q}_{\pm 0.2}|$, \(q_x= \pm q\), $T$ = 80 K, and $\partial T /  \partial x= \Delta T_i / L$ with $\Delta T_i$ ($i=1,2,3$) being the temperature differences used in the experiment and the sample length $L$ = 1.3 mm. $I_0$ and $\tau_{ q}$ are the free parameters, while all others are as defined above.
Using Eq. \ref{eq:fitting} to fit the $A_{{q}}$ values plotted as scatters for both ${\bf q}_{0.2}$ and ${\bf q}_{-0.2}$ in Fig. 4\textbf{d} of the main text, we obtain the solid lines as the best least-squares linear fitting result (again, refer to Fig. 4\textbf{d}). This solution returns a magnon relaxation time $\tau_{ q}$ = 58 $\pm$ 4 ns in YIG. To put this value into context, we attach a table summarizing the magnon relaxation times reported for YIG close to the Brillouin zone center (see Extended Data Table 1). To the best of our knowledge, only the relaxation time reported in the first row of Extended Data Table 1, 2–60 $\mu$s, has actually been experimentally measured. 

\textcolor{black}{The relaxation time of magnons as a function of momentum $\tau_{\bf q}$ in the entire Brillouin zone is the big unknown that governs the thermal properties of magnetic insulators. To date, it cannot be computed accurately or measured by other techniques. Knowing $\tau_{\bf q}$ allows us to compute essential quantities such as the momentum-specific magnon thermal conductivity and the intrinsic spin Seebeck coefficient\cite{chemical_potential,Long_lifetime} without free parameters. Here we performed such calculations as a reference; however at present it is impossible to compare these momentum-dependent quantities against experimentally extracted momentum-integrated values, due to the limitations of existing models in capturing magnon scattering at higher energies and the lack of experimental data on
 $\tau_{\bf q}$ as a function of ${\bf q}$\cite{zhang_2012PRL,rezende2014magnon,chemical_potential}. } 

\subsection*{Magnon thermal transport at 80 K}

The heat {\bf J}$_\text{H}$ and magnon spin  {\bf  J}$_\text{S}$ current densities can be defined by Eq. \ref{heat_current} and \ref{spin_current} using the magnon distribution $f(\bf{q})$ obtained from Eq. \ref{fq_dis}\cite{Adachi_2013,zhang_2012PRL,REZENDE2016171,rezende2014magnon}. The energy and momentum resolved heat {\bf J}$_{\text{H}x}(\epsilon)$ and magnon spin   {\bf J}$_{\text{S}x}(\epsilon)$ current densities are defined as follows: 
\begin{equation} \label{heat_current}
{{J}_{\text{H}x}(\epsilon)}= \int \frac{{d\textbf{q}}}{(2\pi)^3} \hbar \omega({{q}}) {\nu_{x}({\bf q})} f({\bf{q}}) \delta (\epsilon -\hbar\omega(q))  
\end{equation}
\begin{equation} \label{spin_current}
{{ J}_{\text{S}x}(\epsilon)}= \int \frac{{d\textbf{q}}}{(2\pi)^3} \hbar {\nu_{x}({\bf q})} f({\bf{q}}) \delta (\epsilon -\hbar\omega(q)) 
\end{equation}

The temperature gradient is along the $\hat{\mathbf{x}}$ direction, and $\nu_{x}({\bf q})$ is shown below:
\begin{equation} \label{vx}
\nu_{x}({\bf q}) = \nu_{x} (q)\frac{\bf q}{ q}\cdot \hat{\mathbf{x}} = \nu_{q}\cos\theta 
\end{equation}

In spherical coordinates: 
\begin{equation} \label{coordinates}
\int \frac{{d\textbf{q}}}{(2\pi)^3} = \frac{1}{(2\pi)^3} \int_{0}^{\infty} q^2dq  \int_{0}^{2\pi} d\varphi \int_{-1}^{1} d\cos\theta
\end{equation}

Using the value of the magnon relaxation time $\tau_{{q}}$ extracted from the RIXS experiment at ${\bf q}_{0.2}$ (with an energy of $\sim$ 10 meV) $\tau_{{q}}$= 58 ns, and the magnon distribution function described by Eq. \ref{fq_dis}, we obtain:

\begin{equation} \label{heat_current_q}
 \begin{split}
J_{\text{H}x}(\epsilon_{q_{0.2}}) & = \frac{q^2}{(2\pi)^2} \frac{1}{2Dq} \hbar \omega(q) {\nu_{ q}} \Biggl( \frac{2Dq \tau_q}{\hbar} \frac{(\varepsilon_0 + Dq^2)}{k_B T^2} \frac{e^{(\varepsilon_0 + Dq^2)/(k_B T)}}{\left(e^{(\varepsilon_0 + Dq^2)/(k_B T)}-1\right)^2} \frac{\partial T }{\partial x } \Biggr) \int_{-1}^{1}  \cos ^2 \theta d cos\theta\\
& =5.263 \times 10 ^{28} ~\mathrm{s^{-1}m}^{-2}, ~\text{at} ~ {q}=|{\bf q}_{0.2}|, ~\frac{\partial T }{\partial x }=\frac{ \Delta T_2 }{L }
\end{split}
\end{equation}

\begin{equation} \label{pin_current_1}
 \begin{split}
J_{\text{S}x}(\epsilon_{q_{0.2}}) & = \frac{q^2}{(2\pi)^2} \frac{1}{2Dq} \hbar {\nu_{ q}} \Biggl( \frac{2Dq \tau_q}{\hbar} \frac{(\varepsilon_0 + Dq^2)}{k_B T^2} \frac{e^{(\varepsilon_0 + Dq^2)/(k_B T)}}{\left(e^{(\varepsilon_0 + Dq^2)/(k_B T)}-1\right)^2} \frac{\partial T }{\partial x } \Biggr)\int_{-1}^{1}  \cos ^2 \theta d cos\theta\\
& = 3.441 \times 10^{15}~\mathrm{m}^{-2}, ~\text{at}~ {q}=|{\bf q}_{0.2}|, ~\frac{\partial T }{\partial x }=\frac{ \Delta T_2 }{L }
\end{split}
\end{equation}

From these values, we can calculate the momentum resolved magnon thermal conductivity $K_m ({\epsilon_q}) = {{ J}_{\text{H}x}}({\epsilon_q}) / \frac{\partial T }{\partial x }$
and the intrinsic spin Seebeck coefficient $S_m ({\epsilon_q})= 2eT {{ J}_{\text{S}x}}({\epsilon_q})/{(\frac{\partial T }{\partial x } \hbar) }$, considering $\frac{\partial T }{\partial x }$ = $\Delta T_2/L$ = 6385 K/m ($\Delta T_2$ = 8.3~K as in main text, 
 sample length $L$ = 1.3 mm). 
Therefore, the momentum resolved thermal conductivity $K_m{({\epsilon_{q_{0.2}}})} = {{J}_{\text{H}x}}({\epsilon _{q_{0.2}}}) / \frac{\partial T }{\partial x }=8.244 \times 10^{24}$~\text{s}$^{-1}$\text{m}$^{-1}$\text{K}$^{-1}$ and spin Seebeck coefficient $S_m{({\epsilon_{q_{0.2}}})} = 2eT{{ J}_{\text{S}x}}({\epsilon_{q_{0.2}}}) / {(\frac{\partial T }{\partial x } \hbar)} = 1.311  \times 10^{29}$~\text{A}\text{m}$^{-1}$\text{J}$^{-1}$ at $T$ = 80 K. %
We want to emphasize that the presented values are calculated based on a single spherical energy surface \textcolor{black}{corresponding to $\epsilon_{\bf{q}_{0.2}}$}, \textcolor{black}{noting that energy and momentum resolved quantities cannot be directly compared with the integrated observables without additional model assumptions} \cite{chemical_potential,Boona_us,rezende2015thermal}. Ideally, by measuring the magnon spin current as a function of momentum across the entire Brillouin zone, we could obtain estimates for the thermal conductivity $K_m$  and intrinsic spin Seebeck coefficient $S_m$. While the RIXS relaxation time at ${\bf q}_{0.2}$ agrees with published numbers for $q \sim 0$, it does not reflect the fast decay of spin waves as a function of energy as inferred from magnon transport studies \cite{Boona_us,rezende2015thermal,rezende2014thermal,ratkovski2020thermal}. This unsolved issue requires more research, particularly in experimentally extracting the momentum dependence of the magnon relaxation time.


\section*{Acknowledgements}
This work was primarily supported by the US Department of Energy (DOE) Office of Science, Early Career Research Program. The SSE setup was co-supported as part of Programmable Quantum Materials, an Energy Frontier Research Center funded by the U.S. Department of Energy (DOE), Office of Science, Basic Energy Sciences (BES), under award DE-SC0019443. This research used beamline 2-ID of the National Synchrotron Light Source II, and the Nanofabrication and Electron Microscopy facilities of the Center for Functional Nanomaterials (CFN), which are US DOE Office of Science User Facilities operated for the DOE Office of Science by Brookhaven National Laboratory under contract no. DE-SC0012704. 
TK and ES are supported by JST CREST (JPMJCR20C1 and JPMJCR20T2), Grant-in-Aid for Scientific Research (Grants No. JP19H05600 and JP24K01326), and Grant-in-Aid for Transformative Research Areas (Grant No. JP22H05114) from JSPS KAKENHI, MEXT Initiative to Establish Next-generation Novel Integrated Circuits Centers (X-NICS) (Grant No. JPJ011438), Japan, and the Institute for AI and Beyond of the University of Tokyo. Finally, we are grateful to P. Gambardella, J.~P. Hill, and C. Mazzoli for fruitful discussions, and to D. Bacescu for the engineering support with the sample environment.

\section*{Author contributions}

VB conceived the research project. YG and VB designed the study, with contributions from GB, JB, and JP. JJB checked the magnetization measurement of the YIG single crystal with guidance from CR. YG fabricated the SSE devices with help from VB, FC, KK, TK, ES, DNB, and JS. YG, JL, JP, and VB carried out the RIXS measurements and performed the first data interpretation. YG analyzed the RIXS data with guidance from VB. JB and GB contributed to the data discussion and the linearized Boltzmann model. LL tested finite element analysis of the temperature gradient across the sample. YG and VB wrote the manuscript with contributions from all authors.

\section*{Data availability}
Data supporting this study's findings are available upon reasonable request from the corresponding authors.

\section*{Competing interests}
The authors declare no competing interests.

\newpage
\setcounter{figure}{0}
\renewcommand{\figurename}{\textbf{Extended Data Figure}}
\renewcommand{\tablename}{\textbf{Extended Data Table}}

\begin{table}[!h]
\begin{center}
\begin{tabular}{ c } 
\centering
\includegraphics[width=0.4\linewidth]{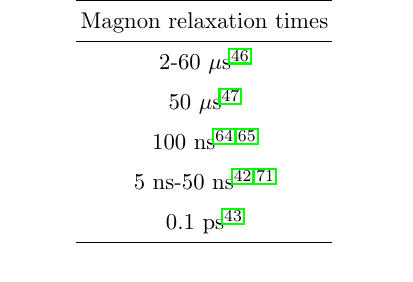}
\end{tabular}
\caption{Reported magnon relaxation times at the Brillouin zone center.}\label{tab:Relaxation time}
\end{center}
\end{table}

\begin{figure}[!h]
\centering	
\centering
\includegraphics[width=0.95\linewidth]{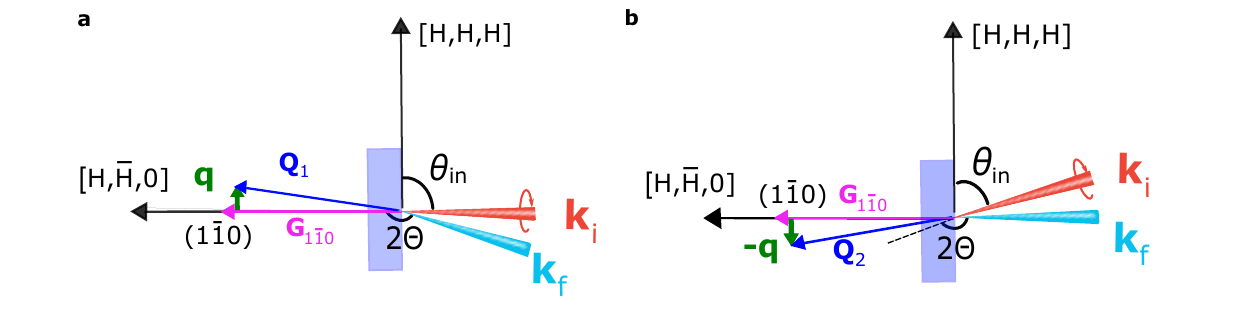}
\caption{\textbf{Top view of the RIXS scattering geometry}. \textbf{a,} Experimental set-up for momentum transfer $\mathbf{q}_{0.2}$ = [0.2, 0.2, 0.2] r.l.u.. The circularly polarized x-rays {{\bf k}$_\text{i}$} impinge the sample at an angle $\theta_{\text{in}} = 90^{\circ}$. The scattering angle $2\Theta$  between incoming beam {{\bf k}$_\text{i}$} and outgoing beam {{\bf k}$_\text{f}$} is fixed at 150$^{\circ}$. \textbf{b,} As \textbf{a} but for momentum transfer $\mathbf{q}_{-0.2}$ = [-0.2, -0.2, -0.2] r.l.u., with $\theta_{\text{in}} = 60^{\circ}$ and $2 \Theta=150^{\circ}$.
\label{fig:momentum}}
\end{figure}

\begin{figure}[!h]
\centering	
\centering
\includegraphics[scale=0.48]{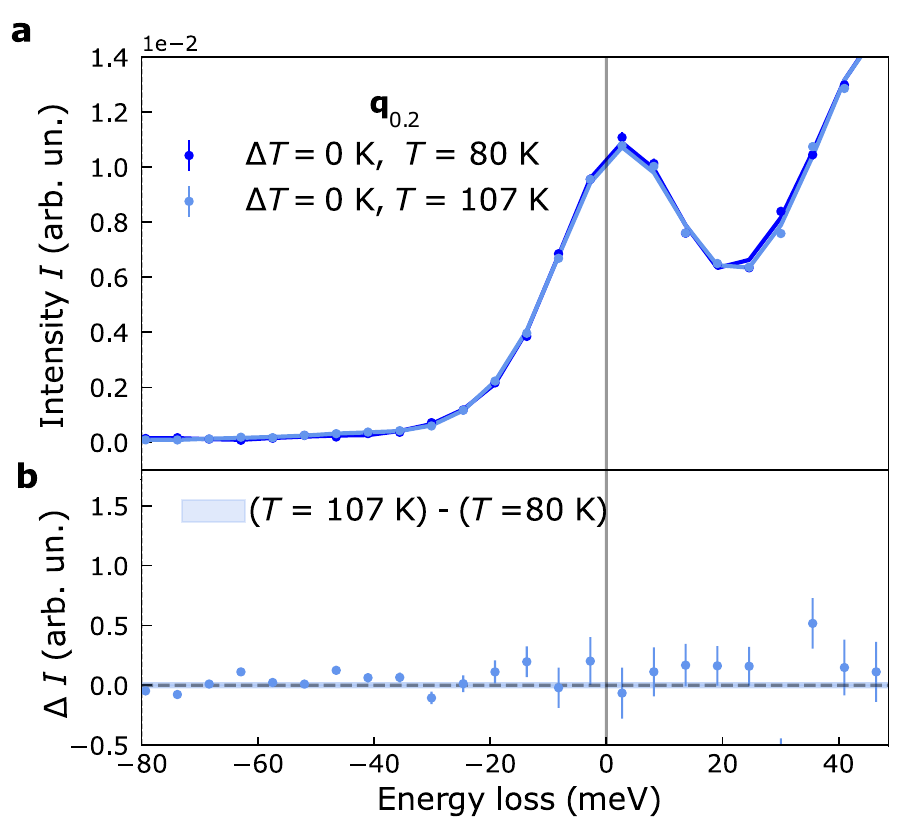}
\caption{\textbf{RIXS measurements as a function of temperature on a reference sample.} \textbf{a,} RIXS spectra measured for ${\bf q}_{0.2}$with $T$ = 80 K (blue scattered line) and $T$ = 107 K (light blue scatterred line) from the reference sample. \textbf{b,} The intensity difference $\Delta I$ between the spectra shown in the panel \textbf{a} drops into the noise level. \label{fig:ref}}
\end{figure}

\begin{figure}[!h]
\centering	
\centering
\includegraphics[width=\linewidth]{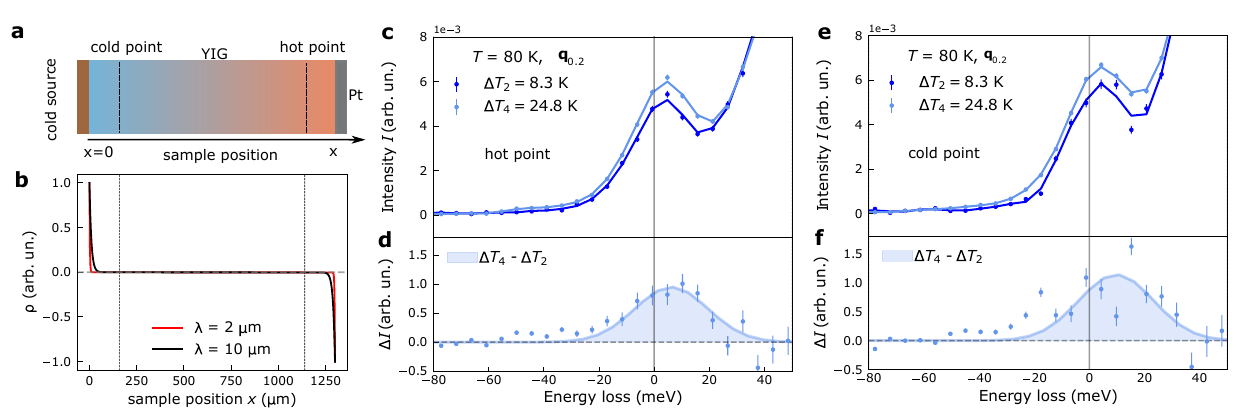}
\caption{\textbf{RIXS spectra at different sample locations.} \textbf{a,} Sketch of the device, in which the color bar indicates temperature differences, similar to Fig. 2\textbf{a} in the main text. \textbf{b,} Magnon accumulation/depletion as a function of position $x$ calculated for a constant gradient in a symmetric sample. The accumulation/depletion decays on the scale of the magnon relaxation length  \(\lambda =2 \mathrm{\mu m} \) in red and 10 $\mathrm{\mu m}$ in black. \textbf{c} (\textbf{e}), RIXS spectra for $\mathbf{q}_{0.2}$ measured at the hot point (cold  point) for two temperature gradients $\Delta T_2$ (dark blue) and $\Delta T_4$ (light blue). \textbf{d} (\textbf{f}), The RIXS intensity difference between spectra for $\Delta T_2$ and $\Delta T_4$ and their Gaussian fit (solid line).\label{fig:pos}}
\end{figure}

\begin{figure}[!h]
\centering	
\centering
\includegraphics[width=1.0\linewidth]{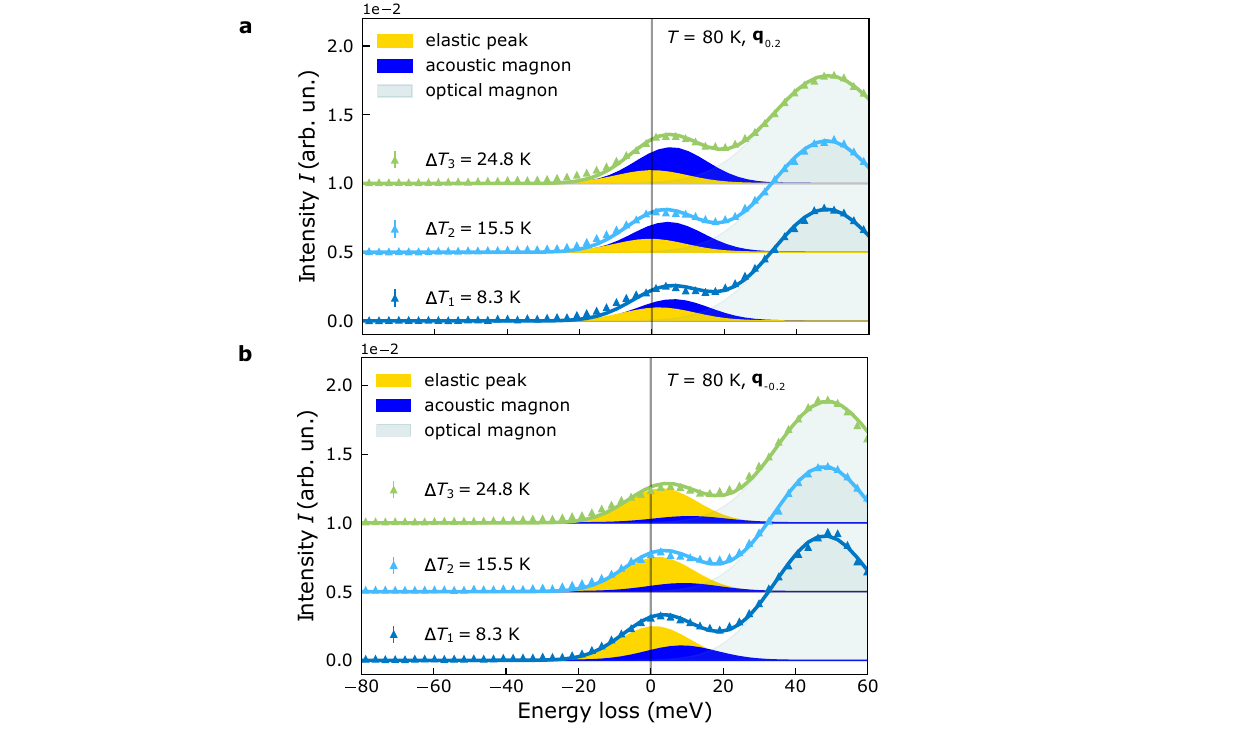}
\caption{\textbf{Gaussian fitting of the RIXS spectra in the non-equilibrium state}. The data were all acquired at $T$ = 80 K and versus $\Delta T$ ($\Delta T_2$, $\Delta T_3$, and $\Delta T_4$),  at ${\bf q}_{0.2}$= [0.2, 0.2, 0.2] r.l.u. \textbf{a,}  and at ${\bf q}_{-0.2}$ = [-0.2, -0.2, -0.2] r.l.u. \textbf{b}. The triangles refer to the experimental data points and the solid lines are the fit by the model in Fig. 1\textbf{c}. The (gold, blue, light-blue)-shaded areas are the extracted (elastic, acoustic magnon, optical magnon)  contributions, respectively.
\label{fig:fitting}}
\end{figure}

\clearpage

\bibliographystyle{naturemag}

\bibliographystyle{naturemag}


\end{document}